\documentclass[prd,aps,11pt,showkeys,nofootinbib,showpacs]{revtex4-1}

\usepackage{amssymb,amsmath}

\usepackage[
    colorlinks,%
    linkcolor=blue,citecolor=red,urlcolor=blue,
]{hyperref}
\usepackage[all,dvips,knot,color]{xy}
\usepackage{tikz}
\usepackage{graphicx}
\usepackage{wrapfig}
\usepackage{calc}

\def\barr{\begin{array}}
\def\earr{\end{array}}
\def\half{\frac{1}{2}}
\def\ben{\begin{equation}}
\def\een{\end{equation}}
\def\bena{\begin{eqnarray}}
\def\eena{\end{eqnarray}}
 
\newcommand{\sect}[1]{\setcounter{equation}{0}\section{#1}}

\def\bN{\mathbb{N}}
\def\bR{\mathbb{R}}
\def\bC{\mathbb{C}}

\def\dd{{\rm d}}
\def\SU{{\rm SU}(2)}
\newcommand{\grqc}[1]{\href{http://arxiv.org/abs/gr-qc/#1}{arXiv:gr-qc/#1}}

\newcommand{\arxiv}[1]{\href{http://arxiv.org/abs/#1/}{arXiv:#1}}

\newcommand{\webpage}[1]{{\color{blue}}\url{#1}{\color{blue}}}

\begin{document}

\title{Emergence of a low spin phase in group field theory condensates}
\author{Steffen Gielen}
\affiliation{Theoretical Physics, Blackett Laboratory, Imperial College London, London SW7 2AZ, U.K. (still EU)}
\email{s.gielen@imperial.ac.uk}

\date{\today}

\begin{abstract}

Recent results have shown how quantum cosmology models can be derived from the effective dynamics of condensate states in group field theory (GFT), where `cosmology is the hydrodynamics of quantum gravity': the classical Friedmann dynamics for homogeneous, isotropic universes, as well as loop quantum cosmology (LQC) corrections to general relativity have been shown to emerge from fundamental quantum gravity. We take one further step towards strengthening the link with LQC and show, in a class of GFT models for gravity coupled to a free massless scalar field and for generic initial conditions, that GFT condensates dynamically reach a low spin phase of many quanta of geometry, in which all but an exponentially small number of quanta are characterised by a single spin $j_0$ (i.e. by a constant volume per quantum). As the low spin regime is reached, GFT condensates expand to exponentially large volumes, and the dynamics of the total volume follows precisely the classical Friedmann equations. This behaviour follows from a single requirement on the couplings in the GFT model under study. We present one particular simple case in which the dominant spin is the lowest one: $j_0=0$ or, if this is excluded, $j_0=1/2$. The type of quantum state usually assumed in the derivation of LQC is hence derived from the quantum dynamics of GFT. These results confirm and extend recent results by Oriti, Sindoni and Wilson-Ewing in the same setting.

\end{abstract}


\maketitle

\sect{Introduction}

Early universe cosmology is widely seen as the most important arena where quantum gravity could have observational consequences, potentially leading to the validation or exclusion of different quantum gravity approaches as well as providing new input for theoretical cosmology. For practitioners of quantum gravity, this poses the challenge of deriving an effective description at cosmologically relevant scales from the fundamental dynamics defined at the Planck scale. For a fundamental description in terms of discrete excitations of quantum geometry, this process also encompasses the emergence of an effective continuum spacetime from a large number of fundamental quanta \cite{emergence}. This situation is familiar in condensed matter physics: phenomena such as superfluidity can be described macroscopically in terms of a continuum hydrodynamic description, but have a microscopic origin in Bose--Einstein condensation in many-particle quantum physics \cite{BECbook}.

In the context of loop quantum gravity (LQG) \cite{LQG} and its second quantisation formulation \cite{2ndquant} {\em group field theory} (GFT) \cite{GFT}, a line of research has been following this analogy, describing a macroscopic Universe as a {\em condensate of quanta of geometry}, themselves given by quantum excitations of a GFT field \cite{GFCreview}. Rather than following the traditional minisuperspace strategy of quantum cosmology \cite{QCreview} and `quantising' large-scale degrees of freedom such as the scale factor, one aims to derive an effective description at large scales from the `hydrodynamics of quantum gravity'. Being able to obtain such an effective description provides an important self-consistency test for quantum gravity, as well as leading to potentially novel and interesting phenomenology to be explored in cosmology. In the context of LQG and GFT, a particular focus has been on understanding the possible link between loop quantum cosmology (LQC), which is derived using minisuperspace techniques together with input from LQG \cite{LQC}, and a more fundamental description, as in GFT. Indeed, the idea of the universe as a condensate-like configuration of many interacting `patches' can also be understood as a general approach for going beyond the minisuperspace truncation \cite{Martinreview}.

One of the initial aims of work on GFT condensates has been to derive the classical Friedmann equations for homogeneous, isotropic universes in an appropriate large-scale or semiclassical limit, starting with \cite{letter,JHEP} where a WKB approximation was employed to this effect. In particular, \cite{JHEP} introduced a natural way of coupling a free, massless scalar field to gravity in the context of GFT, showing that this leads to the correct matter coupling also at the level of the emergent Friedmann equations, in the same WKB limit, with an effective Newton's constant related to couplings in the GFT action. Considering models of gravity coupled to a massless scalar is of particular interest when comparing with LQC where such a scalar appears as a relational clock variable. A more general class of GFT actions with a similar coupling to a massless scalar field was then recently studied in much more detail in \cite{DLElong,DLEshort}, and several interesting results were found that strengthen the link of GFT condensate dynamics to minisuperspace models of LQC: a systematic way of imposing isotropy on the state led to simpler effective (Gross--Pitaevskii-like) equations for the `condensate wavefunction' (or mean field). In a weak-coupling approximation, these could be translated into {\em relational Friedmann equations} for the volume of the universe $V(\phi)$, seen as a function of `relational time' $\phi$ given by the scalar field\footnote{The scalar field can serve as a clock since its momentum $\pi_\phi$ is conserved and therefore $\phi$ evolves monotonically.}. It was shown that, for a generic choice of mean field and at large volume\footnote{Note that large volume corresponds to small curvature when the Friedmann equation holds, which is what one would expect as the semiclassical limit.}, these reduce precisely to the Friedmann equations of classical general relativity coupled to a massless scalar (where the exact solution is $V(\phi)=V_0\exp(\pm\sqrt{12\pi G}\phi)$), provided an infinite number of parameters in the GFT action are equal (and then identified with $3\pi G$); approximate equality corresponds to classical Friedmann equations with corrections. In the special case of a {\em single-spin} condensate where only quantum geometric excitations of a fixed spin $j=j_o$ (i.e. fundamental quanta with given fixed areas or volumes) are excited, only a single parameter needed to be fixed to recover the Friedmann equations, which then also include high-curvature corrections precisely of the form of the effective equations of LQC, plus further quantum corrections. 

The single-spin assumption is particularly relevant for LQC, where in order to derive the quantum-corrected Friedmann equation one uses the picture of a Universe consisting of a large number of LQG quanta with spin $j=1/2$, so that fundamental plaquette areas take the minimum non-zero value in LQG (see, e.g., section IIIB of \cite{BianchiI}); expansion or contraction of the Universe proceeds by creation or annihilation of fundamental quanta. This leads to the idea of {\em lattice refinement} for LQC \cite{latticeref} from which the `improved dynamics' form of holonomy corrections \cite{improdyn} is thought to originate. Some steps towards understanding lattice refinement from the perspective of GFT condensates were taken in \cite{NJP}, leaving the crucial relation $N=N(V)$ between the total number of quanta and the volume of the Universe (or, equivalently, the scale factor) undetermined.

In this paper, we strengthen and extend the results reported in \cite{DLElong,DLEshort}. Namely, within the class of GFT models considered in \cite{DLElong,DLEshort}, we exhibit a single condition -- that the ratio $B_j/A_j$ of $j$-dependent couplings $A_j$ and $B_j$ takes a positive maximum for some $j>0$ -- for the following to occur: if the condition is satisfied, a phase in which almost all quanta in a GFT condensate have the same spin emerges {\em dynamically} in the mean-field approximation, for generic choices of mean field. In this process, the Universe grows to an exponentially large volume and follows precisely the classical Friedmann dynamics. Compared to \cite{DLElong,DLEshort}, these new results show that a large-volume or small-curvature limit in which the Friedmann equations hold is reached generically, with weaker assumptions on the GFT model than considered before; they show that the single-spin assumption is always satisfied asymptotically; the latter result then also leads directly to the `improved dynamics' scheme in LQC. Furthermore, it was already shown in \cite{DLElong,DLEshort} that the volume has a non-zero minimum leading to a bounce and resolution of singularities, and this remains true here. In order to clarify the link to previous papers, we first show our results in the case of the model proposed in \cite{JHEP}, and then generalise to the wider class of models of \cite{DLElong,DLEshort}. In the model of \cite{JHEP} and when there are no quanta with spin $j=0$, a $j=1/2$ phase emerges, giving further support for the model that adds to its motivation from the theoretical GFT perspective.

An important remark concerns a difference between the Hilbert spaces of GFT and LQG \cite{newGFTreview}: in LQG, excitations with $j=0$ have no physical significance; by cylindrical consistency, a state containing $j=0$ links is equivalent to one in which these are removed \cite{LQG}. This is not true in GFT, where $j=0$ excitations contribute to the total number of quanta but (for the most common choices of area and volume operators) have zero volume and area, so that a match between the LQC and GFT dynamics appears impossible for these $j=0$ quanta. We take this fact into account by only considering GFT condensates with no $j=0$ excitations. Within the approximations taken, this condition is preserved by the dynamics, and is therefore a consistent restriction on the choice of mean field, within the mean-field approximation. It constitutes fine-tuning from the perspective of GFT condensates, but appears necessary for a closer link of GFT condensates to LQG.

\sect{Cosmology from GFT condensates}

In this section we review the most essential aspects of GFT as an approach to quantum gravity, the extraction of an effective dynamics for coherent condensate states, and the connection to cosmology. For more background, we refer to the review \cite{GFCreview} and to the extensive discussions in \cite{JHEP, DLElong}.

In GFT, spacetime geometry emerges from excitations of a (complex) quantum field $\varphi$, defined on a group manifold that is {\em not} physical space or spacetime, but the configuration space of the degrees of freedom of a tetrahedron (with boundary). Concretely, if the local gauge group of gravity is taken to be $\SU$, the gravitational configuration space is $\SU^4/\SU$: there are four parallel transports associated to four links dual to the faces of the tetrahedron, with invariance under diagonal right multiplication by an element of $\SU$ which corresponds to a discrete gauge transformation at the vertex where these links meet. In the models we are interested in, one also includes a scalar field $\phi$ as an additional degree of freedom on the vertex of the tetrahedron \cite{JHEP,DLElong}. $\varphi$ is then defined as
\ben
\varphi: \SU^4\times \bR \rightarrow \bC\,,\quad \varphi(g_I,\phi)=\varphi(g_Ih,\phi)\quad\forall h\in\SU\,.
\label{fielddef}
\een
Here and in the following, the index $I$ runs over the four $\SU$ arguments of $\varphi$, corresponding to four faces of the tetrahedron. The domain space of the scalar field is $\bR$; an alternative choice would be ${\rm U}(1)$, the space of `point holonomies' used in LQG \cite{JHEP}.

Dynamics are now defined through an action
\ben
S[\varphi,\bar\varphi]=-\int (\dd g)^4\,\dd\phi\;\bar\varphi(g_I,\phi)\mathcal{K}\varphi(g_I,\phi) + \mathcal{V}[\varphi,\bar\varphi]
\label{akshn}
\een
which we split into kinetic (quadratic) and interaction parts, where the kinetic part is assumed to be local. A trivial kinetic term ($\mathcal{K}=1$) is sufficient to define a class of GFT models whose Feynman amplitudes correspond to the amplitudes of a class of spin foam models \cite{RovelliReis} through the choice of $\mathcal{V}$. This indeed motivated GFT as a field-theoretic description of spin foam dynamics. More general choices of $\mathcal{K}$ are possible and, as usual, renormalisation affects the form of the action and can generate additional terms. In particular, a Laplace--Beltrami operator on $\SU^4$ seems naturally generated by quantum fluctuations \cite{radiative} (see also e.g. \cite{GFTrenorm} for recent more general results on GFT renormalisation), motivating us to study such more general choices of $\mathcal{K}$.

The theory is now defined either through a covariant path integral (with perturbative expansion in Feynman amplitudes) or in an operator formalism. In the latter, one promotes $\varphi$ and $\bar\varphi$ to field operators with bosonic commutation relations
\ben
[\hat\varphi(g_I,\phi),\hat\varphi(g'_I,\phi')]=[\hat\varphi^\dagger(g_I,\phi),\hat\varphi^\dagger(g'_I,\phi')]=0\,,\quad [\hat\varphi(g_I,\phi),\hat\varphi^\dagger(g'_I,\phi')]={\bf 1}(g_I,g'_I)\delta(\phi-\phi')\,,
\een
where ${\bf 1}(g_I,g'_I)=\int \dd h\prod_{I=1}^4 \delta(g_I h (g'_I)^{-1})$ is a delta distribution on $\SU^4/\SU$. One can then define a Fock space, starting from a `no-space' vacuum $|0\rangle$ annihilated by $\hat\varphi(g_I,\phi)$;
\ben
\hat\varphi^\dagger(g_I,\phi)|0\rangle \equiv |g_I,\phi\rangle
\een
is a one-particle state, representing a single tetrahedron with connection data given by parallel transports $g_I$ and the value $\phi$ for the scalar field. Such a single-particle state is just the smallest perturbation of a vacuum in which no space is present at all; in order to describe a macroscopic continuum Universe one needs to excite a large number of GFT `particles'. A good candidate for this is a GFT {\em condensate} state, whose main property is {\em wavefunction homogeneity}: the state is determined by a single-particle wavefunction. 

The simplest choice for such a state is
\ben
|\sigma\rangle =\mathcal{N}(\sigma) \exp\left(\int (\dd g)^4\,\dd\phi\;\sigma(g_I,\phi)\,\hat\varphi^\dagger(g_I,\phi)\right)|0\rangle\,,
\label{condendef}
\een
where $\sigma(g_I,\phi)$ is called the `condensate wavefunction' and $\mathcal{N}(\sigma)$ is a normalisation factor. $\sigma(g_I,\phi)$ itself is not normalised, but determines the number of quanta in the state;
\ben
N(\phi)=\langle\sigma|\hat{N}(\phi)|\sigma\rangle\equiv \int (\dd g)^4 \;\langle\sigma|\hat\varphi^\dagger(g_I,\phi)\hat\varphi(g_I,\phi)|\sigma\rangle = \int (\dd g)^4\;|\sigma(g_I,\phi)|^2
\een
is the expectation value of the number of quanta $N(\phi)$ at the value $\phi$ of the scalar field \cite{DLElong}. For further discussion of the role of wavefunction homogeneity, arguments for the interpretation of $|\sigma\rangle$ as a macroscopic, homogeneous geometry, and the interpretation of $\sigma$ as a wavefunction, see \cite{GFCreview}. One can define more general condensates satisfying wavefunction homogeneity, containing additional topological structure and defined by a sum over connected graphs of arbitrary complexity \cite{gencond}. 

Due to (\ref{fielddef}), $\sigma$ satisfies $\sigma(g_I,\phi)=\sigma(g_Ih,\phi)$ for all $h\in\SU$; in order for the state to only contain the gauge-invariant degrees of freedom of a tetrahedron, one also imposes left invariance $\sigma(g_I,\phi)=\sigma(kg_I,\phi)$ for all $k\in\SU$, so that $\sigma$ is a function on $\SU\backslash\SU^4/\SU$, which is isomorphic to the space of connection degrees of freedom of a homogeneous Universe \cite{CQG}.

Equation (\ref{condendef}) is a coherent state, satisfying
\ben
\hat\varphi(g_I,\phi)|\sigma\rangle = \sigma(g_I,\phi)|\sigma\rangle
\een
which implements the {\em mean-field approximation}: all normal-ordered expectation values of products of field operators can be evaluated by replacing $\hat\varphi$ by $\sigma$ and $\hat\varphi^\dagger$ by $\bar\sigma$. This approximation is commonly used in condensed matter physics, where it captures well the physics of diluted, weakly interacting Bose--Einstein condensates \cite{BECbook}. It represents a semiclassical limit, in which field operators can be replaced by classical fields. It is thus very well suited for describing the emergence of a semiclassical, macroscopic and nearly homogeneous universe in which spatial gradients are small. Just as in condensed matter physics, one expects the mean-field approximation to break down when interactions get stronger, i.e. in the early Universe in cosmology. We refer to \cite[section 4.2]{DLElong} for a discussion of criteria for the validity of the mean-field approximation in the GFT context.

Within the mean-field approximation, equations encoding the GFT quantum dynamics can be rewritten as conditions on the mean field $\sigma(g_I,\phi)$. As a first approximation to the full dynamics, consider the expectation value of the GFT operator equation of motion,
\ben
\langle\sigma|\frac{\delta\hat{S}[\hat\varphi,\hat\varphi^\dagger]}{\delta\hat\varphi^\dagger(g_I,\phi)}|\sigma\rangle = \frac{\delta S[\sigma,\bar\sigma]}{\delta\bar\sigma(g_I,\phi)}=-\mathcal{K}\sigma(g_I,\phi) + \frac{\delta \mathcal{V}[\sigma,\bar\sigma]}{\delta\bar\sigma(g_I,\phi)}=0\,.
\label{grosspit}
\een
In condensed matter physics, this procedure leads to the {\em Gross--Pitaevskii equation} for the condensate wavefunction. For a GFT condensate, (\ref{grosspit}) and corresponding higher order equations define the cosmological dynamics. Indeed, one of the main goals of work on GFT condensates is to show how (\ref{grosspit}) leads to effective cosmological dynamics consistent with general relativity at low curvature, and with interesting quantum corrections.

\sect{Solving the condensate dynamics}

Let us start by revisiting the GFT for gravity coupled to a scalar field introduced in \cite{JHEP}, with
\ben
\mathcal{K}=\mu+ \sum_I \Delta_{g_I} + \tau\frac{\partial^2}{\partial\phi^2}\,,
\label{kinetic}
\een
where $\Delta_{g_I}$ is the Laplace--Beltrami operator acting on the $I$th $\SU$ argument, and we have absorbed a possible coupling in front of it into a normalisation of the action. As we have mentioned, adding a Laplacian in $g_I$ to a constant $\mathcal{K}$ is well motivated by studies of GFT renormalisation, and it is natural to include the same derivative with respect to the matter variable $\phi$. The latter, in turn, has also been derived from independent arguments as a way to couple a free scalar field to gravity in the GFT setting \cite{DLElong,DLEshort}.  

Using (\ref{kinetic}) means that one can study a weak-coupling limit in which the contribution of $\mathcal{V}$ in (\ref{grosspit}) is neglected. This approximation has been employed in most studies so far \cite{GFCreview}, and is again motivated by the necessity for the Universe to be nearly homogeneous with small spatial gradients. From the microscopic point of view, interactions have to be weak for the simple coherent states (\ref{condendef}) to be a reliable approximation to the exact physical state.

The resulting equation of motion for $\sigma(g_I,\phi)$,
\ben
-\left(\mu+ \sum_I \Delta_{g_I} + \tau\frac{\partial^2}{\partial\phi^2}\right)\sigma(g_I,\phi) = 0\,,
\label{dyneq}
\een
now admits an easy direct solution. Using the Peter--Weyl decomposition \cite{PeterWeyl} of functions on $\SU$ into representations, together with $\sigma(g_I,\phi)=\sigma(g_Ih,\phi)=\sigma(kg_I,\phi)$, gives \cite[equation (52)]{DLElong}
\ben
\sigma(g_I,\phi)= \sum_{j_I,\iota_l,\iota_r}\sigma^{j_I\iota_l\iota_r}(\phi)\;C^{j_I\iota_l\iota_r}(g_I)
\een
where $j_I\in\bN_0/2$ are representation labels (spins) for $\SU$, and $C^{j_I\iota_l\iota_r}(g_I)$ are given, up to factors of $(2j_K+1)$, by a product of four Wigner $D$-matrices $D^{j_K}_{m_K n_K}(g_K)$ (for $K=1,\ldots,4$) with their $m$ and $n$ indices contracted by intertwiners $\iota_l$ and $\iota_r$. The Wigner matrices are eigenfunctions of the $\SU$ Laplacian, so that (\ref{dyneq}) becomes
\ben
\left(\sum_I j_I(j_I+1) -\mu - \tau\frac{\partial^2}{\partial\phi^2}\right)\sigma^{j_I\iota_l\iota_r}(\phi) = 0\,.
\label{jdyneq}
\een
As shown in \cite{DLElong}, a natural way to impose isotropy on the condensate is to demand that it consists of equilateral tetrahedra: $\sigma^{j_I\iota_l\iota_r}(\phi)$ is only non-zero if all $j_I$ are equal, and the intertwiners $\iota_l$ and $\iota_r$ are fixed to maximise the tetrahedron volume for given $j$. With these assumptions, the first term in the brackets in (\ref{jdyneq}) becomes $4j(j+1)$, and the general solution is simply
\ben
\sigma_j(\phi)=\alpha^+_j\exp\left(\sqrt{\frac{4j(j+1)-\mu}{\tau}}\,\phi\right)+\alpha^-_j\exp\left(-\sqrt{\frac{4j(j+1)-\mu}{\tau}}\,\phi\right)\,.
\label{exactsol}
\een
The behaviour of these solutions clearly depends on whether $(4j(j+1)-\mu)/\tau$ is positive or negative: they either exponentially grow and decay or oscillate. We assume $\tau<0$, which is required for the correct coupling of matter to gravity in the effective Friedmann equation \cite{JHEP}. Then, for $\mu>0$, the behaviour of the different spin components changes dramatically above a critical spin $j_c=\half \lfloor \sqrt{\mu+1}-1\rfloor$; the $\sigma_j$ with $j>j_c$ oscillate around $\alpha_j^\pm$, whereas $\sigma_j$ with $j\le j_c$ generically (unless some $\alpha^\pm_j$ are zero) grow exponentially as $\phi\rightarrow\pm \infty$, with fastest growth for $j=0$, followed by $j=\half$, etc. We exclude $\mu<0$, where there are only oscillating solutions which do not lead to a macroscopic universe.

As $|\sigma_j(\phi)|^2$ is the expectation value for the number of quanta in the $j$ representation at `time' $\phi$, the condensate is quickly dominated by $j=0$ quanta, with occupation numbers for higher $j$ exponentially suppressed. As we have said, such quanta do not have a clear interpretation from the perspective of LQG, where one could remove them using cylindrical consistency. In order to facilitate comparison between the effective GFT dynamics and LQC, we hence set $\alpha_0=\beta_0=0$. This is a finely-tuned initial condition in GFT, but self-consistent as all $j$ are decoupled in (\ref{jdyneq}).

The dominant contribution at large $|\phi|$ then comes from $\sigma_{1/2}(\phi)$, and the total volume $V(\phi)$ at the value $\phi$ of the scalar field asymptotes to
\ben
V(\phi) = \sum_j V_j\,|\sigma_j(\phi)|^2 \;\stackrel{\phi\rightarrow\pm\infty}{\sim}\; V_{1/2} \left|\alpha^\pm_{1/2}\right|^2 \exp\left(\pm 2\sqrt{\frac{\mu-3}{|\tau|}}\,\phi\right)
\label{volumeev}
\een
with the expression for $V(\phi)$ given in \cite{DLElong}; $V_{1/2}$ is the (Planckian) volume in LQG for an equilateral tetrahedron with $j=1/2$ (and more generally, $V_j\sim j^{3/2}V_{1/2}$). For positive and negative large $\phi$, (\ref{volumeev}) reduces to the solution for the volume of a homogeneous, isotropic Universe in classical general relativity coupled to a massless scalar field
\ben
V=V_0\exp(\pm\sqrt{12\pi G}\phi)\,,
\een
if we identify $\frac{\mu-3}{|\tau|}=3\pi G$. In particular, (\ref{volumeev}) in this regime satisfies the classical Friedmann equations. Due to the exponentials in (\ref{exactsol}), this regime is reached quickly as $\phi\gtrsim m_{{\rm Pl}}$, with emergent Planck mass $m_{{\rm Pl}}\sim G^{-1/2}=\sqrt{\frac{3\pi|\tau|}{\mu-3}}$. The full solution $V(\phi)$ interpolates between a contracting and an expanding classical Universe, reminiscent of the bounce in LQC \cite{LQC}.

Note that the last term in (\ref{volumeev}) is proportional to the asymptotic total number of quanta $N(\phi)$ and thus for $\phi\gtrsim m_{{\rm Pl}}$, $N(V)\propto V$, the relation needed for the improved dynamics scheme \cite{improdyn} in LQC. This is merely a different way of saying that asymptotically almost all quanta have spin $j=1/2$, fixing the ratio of total volume to the number of quanta.

Using methods developed in \cite{DLElong,DLEshort} for relational observables in GFT and excluding GFT quanta with $j=0$, we have thus shown that the GFT kinetic term (\ref{kinetic}) with $\mu>0$ and $\tau<0$ gives effective dynamics in which generic initial conditions for $j\neq 0$ lead to a regime where almost all quanta have spin $1/2$, and the total volume follows precisely the classical Friedmann equations. This provides strong evidence that this class of GFT models leads to the correct large-scale, semiclassical limit, adding to the motivation for (\ref{kinetic}) in particular from GFT renormalisation. It also strengthens the results in \cite{JHEP} for the emergence of a Friedmann equation with correct coupling to matter.

These results can be strengthened further by straightforward extension to a more general class of GFT models. In particular, one can study more general dynamical equations
\ben
\left(-B_j+ A_j\frac{\partial^2}{\partial\phi^2}\right)\sigma_j(\phi) = 0\,,
\label{generaleq}
\een
as done in \cite{DLElong,DLEshort}, with general solution
\ben
\sigma_j(\phi)=\alpha^+_j\exp\left(\sqrt{\frac{B_j}{A_j}}\,\phi\right)+\alpha^-_j\exp\left(-\sqrt{\frac{B_j}{A_j}}\,\phi\right)\,.
\label{anothersol}
\een
This solution was essentially given in \cite{MarcoMairi}, and used to study the bounce for single-spin condensates. Again, its properties depend on the sign of $B_j/A_j$; $B_j/A_j>0$ leads to exponentially expanding and contracting universes, whereas for $B_j/A_j<0$ the number of quanta $|\sigma_j(\phi)|^2$ oscillates around some initial configuration. Furthermore, for {\em all models} for which $B_j/A_j$ has a positive maximum for some $j=j_0$ (again excluding $j=0$)
\ben
V(\phi) = \sum_j V_j\,|\sigma_j(\phi)|^2 \;\stackrel{\phi\rightarrow\pm\infty}{\sim}\; V_{j_0} \left|\alpha^\pm_{j_0}\right|^2 \exp\left(\pm 2\sqrt{\frac{B_{j_0}}{A_{j_0}}}\,\phi\right)\,;
\een
in this case, to obtain an exponentially expanding Universe satisfying the classical Friedmann dynamics at large volume, for generic choices of mean field, only a single identification $B_{j_0}/A_{j_0}=3\pi G$ is needed. If the maximum occurs at a low $j_0$, a low spin regime emerges dynamically. These observations confirm and extend the results in \cite{DLElong,DLEshort}, where $B_j/A_j=3\pi G$ for all $j$ was assumed to recover the correct classical limit; this is not necessary.

To facilitate further comparison, we compute the conserved quantities $E_j$ and $Q_j$ identified in \cite{DLElong,DLEshort} for the solution (\ref{anothersol}). For $B_j/A_j>0$, they are
\ben
E_j=-4\frac{B_j}{A_j}\,{\rm Re}(\alpha_j^+ \overline{\alpha_j^-})\,,\quad Q_j=2\sqrt{\frac{B_j}{A_j}}\,{\rm Im}(\alpha_j^+ \overline{\alpha_j^-})\,,
\label{EQ1}
\een
whereas for $B_j/A_j<0$
\ben
E_j=-2\frac{B_j}{A_j}(|\alpha_j^+|^2+|\alpha_j^-|^2)\,,\quad Q_j = \sqrt{\left|\frac{B_j}{A_j}\right|}(|\alpha_j^+|^2-|\alpha_j^-|^2)\,.
\label{EQ2}
\een
The classical limit assumed in \cite{DLElong,DLEshort} to recover the Friedmann equations is $|\sigma_j|^2\gg \left|\frac{A_j}{B_j}E_j\right|$ and $|\sigma_j|^2\gg \left|\sqrt{\left|\frac{A_j}{B_j}\right|}Q_j\right|$. For $B_j/A_j>0$, this classical limit is reached at $|\phi| \gtrsim \sqrt{\frac{A_j}{B_j}}$ for generic initial conditions, whereas spins with $B_j/A_j<0$ never reach such a regime, as these modes fail to expand sufficiently. The requirement of reaching a classical limit is only a restriction on the values of the couplings in the GFT action, and thus allows distinguishing between different GFT models. For appropriate couplings, the classical limit is reached generically without further assumptions.

The last statement might look puzzling from the perspective of quantum cosmology, where only special states lead to the emergence of a classical Universe; how can it be that this happens generically for GFT condensates? The answer is that the mean-field approximation already is a statement about semiclassicality; failure of semiclassicality is manifest in the breakdown of this approximation, rather than any condition on the `condensate wavefunction', consistent with the conceptual basis of GFT condensate cosmology \cite{CQG,NJP,JHEPalone,GFCreview}. One needs to check whether the assumption of negligible contributions from the GFT interactions that is used in the mean-field approximation is self-consistent \cite{DLElong}.

As a last comment, we repeat the results of \cite{DLElong,DLEshort} regarding the resolution of singularities: as long as one of the $Q_j$ is non-zero, the total volume has a non-zero bound from below. The explicit expressions (\ref{EQ1}) and (\ref{EQ2}) make clear that this condition is satisfied for almost all condensate states, meaning that singularity resolution is generic.

\sect{Conclusions}

Starting from the GFT analogue of the Gross--Pitaevskii equation, defining the effective dynamics of GFT condensates (\ref{condendef}), together with a weak-coupling approximation common in this approach, we have obtained strong results that give further support for the use of condensates for the extraction of cosmological dynamics from GFT. We have confirmed and extended the results of \cite{DLElong,DLEshort} to a wider class of GFT models: if, in (\ref{generaleq}), the ratios $B_j/A_j$ of GFT couplings take a positive maximum for some $j=j_0\neq 0$, consistency with the classical Friedmann equations at large volumes follows directly from setting $B_{j_0}/A_{j_0}=3\pi G$. For such GFT models, a generic condensate will dynamically expand into a configuration in which almost all quanta have spin $j_0$, with other spins exponentially suppressed. This shows explicitly how the requirement on GFT condensates to expand to a large Universe can constrain the possible couplings in the fundamental GFT.

We have shown one example, discussed already in \cite{JHEP}, which is of this form (after we exclude $j=0$ quanta which have no clear interpretation in LQG) and leads to $j_0=1/2$, precisely what is needed in order to derive the intuitive picture used in LQC, and to give a new and convincing derivation of the improved dynamics form \cite{BianchiI,improdyn} of LQC dynamics from a fundamental theory (to be compared with previous derivations from GFT condensates \cite{JHEPalone,Calcagni} and in other LQG settings \cite{AlesciCianfrani}). This result is even more interesting as the kinetic term in this model (\ref{kinetic}) is simple and very natural, given that a Laplace--Beltrami operator on the group appears in GFT renormalisation. The existence of $j=0$ quanta in GFT is indeed one of the main differences to standard LQG \cite{newGFTreview}.

We have not made any statements about the precise form of quantum corrections away from the large-volume regime, where subleading terms corresponding to different $j$ contribute significantly; such corrections depend on the state and on additional couplings that are not fixed by the classical limit. This observation opens the door for a rich phenomenology of GFT condensate cosmology, and the derivation of novel quantum gravity corrections to the classical Friedmann equations, to be explored in future work. Additional assumptions appear necessary to obtain the precise form of the modified Friedmann equations in LQC at high curvature (for example, assuming that only a single spin $j$ is excited throughout \cite{DLElong,DLEshort}), and it needs to be understood better when these hold.

In order to further strengthen the results presented here, one should ultimately go beyond the weak-coupling approximation in (\ref{jdyneq}). Certain forms of the GFT interaction, for example the EPRL \cite{EPRL} interaction considered in \cite{DLElong}, lead to equations that are local in $j$ for isotropic states, so that the decoupling of different $\sigma_j$ modes is maintained. One could hope to adapt methods used in condensed matter physics, such as the Thomas--Fermi approximation for Bose--Einstein condensates \cite{ThomFerm}, to such models, but this is again left for future work.

\begin{acknowledgments}

I would like to thank Daniele Oriti, Edward Wilson-Ewing and an anonymous referee for helpful comments on the manuscript. The research leading to these results has received funding from the People Programme (Marie Curie Actions) of the European Union's Seventh Framework Programme (FP7/2007-2013) under REA grant agreement n$^{{\rm o}}$ 622339.

\end{acknowledgments}

\end{document}